# Automated segmentation of rheumatoid arthritis immunohistochemistry stained synovial tissue

Amaya Gallagher-Syed[1,2], Abbas Khan[2], Felice Rivellese[3], Costantino Pitzalis[3], Myles J. Lewis[3], Gregory Slabaugh[2], Michael R. Barnes[1,2]

[1]Centre for Translational Bioinformatics, Queen Mary University of London
[2]Digital Environment Research Institute, Queen Mary University of London
[3]Centre for Experimental Medicine and Rheumatology, Queen Mary University of London

## 1 Introduction

**Rheumatoid Arthritis** Rheumatoid Arthritis (RA) is a chronic, autoimmune disease of unknown aetiology, affecting circa 1% of the total population. The joint synovial membrane (or synovium) is the major target of RA, which is characterised by synovial inflammation and hyperplasia [9]. The methods available for the study of synovial tissue have advanced considerably in recent years, from arthroplasty and blind needle biopsy to the use of arthroscopic and ultrasonographic technologies which improve the reliability and quality of synovial biopsies [12]. This has led to rapid progress in the study of disease pathogenesis and patient stratification, with increasingly complex analytical and technological approaches [12].

**Digital Pathology** One such approach is the histopathological assessment of joint synovium samples using digital Whole Slide Images (WSIs). The analysis of WSIs can lead to patient diagnosis and treatment by enabling the identification and quantification of spatial organisation and cellular features within the joint [3, 12, 14]. Complementary information can be gathered using several stain types, such as Haematoxylin & Eosin (H&E) and Immunohistochemistry (IHC). IHC in particular stains cellular proteins using specialised antibodies and is therefore well suited to highlighting functional organisation [13].

**Tissue segmentation** Yet much of the pre-processing and analysis of these samples is performed manually and semi-quantitatively by expert pathologists, a labour and knowledge-intensive task which precludes wider access, implementation in clinical practice and research reproducibility [14, 9, 8]. The effectiveness of other widely used medical image segmentation methods such as edge-based techniques [17, 11], active contours [6, 19] or watersheds [5, 4, 10] is limited by the great heterogeneity in stain intensity and colour, the fragmented nature of synovial tissue samples, as well as the presence of many undesirable artefacts present in the WSIs, such as water droplets, pen annotation, folded tissue, blurriness (see Figure 1C for reference) [5]. Furthermore, slides are typically stained with three or more different IHC stains and can originate from a variety of clinical centres, each with their own staining protocol, microscope and digital scanners [8]. There is therefore a need for a robust, automated segmentation algorithm which can cope with this variability.

**Contribution** We provide a fully trained UNet segmentation tool for WSI IHC synovial tissue which can be used as the first step in an automated image analysis pipeline. It is robust to common WSIs artefacts, clinical centre/scanner batch effect and can be used on different types of IHC stains. It can be used as is, or fine-tuned on any IHC musculoskeletal dataset, removing the need for manual segmentation by pathologists and offering a solution to the current image analysis bottleneck. The code is available: `https://github.com/AmayaGS/IHC_Synovium_Segmentation`

## 2  Methods

**Data collection** A total of 164 patients, fulfilling the 2010 American College of Rheumatology/European Alliance of Associations for Rheumatology (EULAR) classification criteria for RA were recruited to the R4RA clinical trial from 20 European centers [15] [7]. Patients underwent ultrasound-guided synovial biopsy of a clinically active joint.

Briefly, samples were then fixed in formalin, embedded in paraffin, cut with microtome and stained with the relevant IHC stains: IHC CD20 (B cells), IHC CD68 (macrophages) and IHC CD138 (plasma cells) [7]. Samples were then placed on glass slides and scanned into Whole Slide Image (.ndpi format) with digital scanners under 40x or 20x objectives.

**Hand labelling** 465 IHC WSIs were manually labelled and full-scale binary Ground Truth (GT) masks were extracted using the QuPath software [2], as shown in Figure 1B.

**Training set** Patient IDs were used to randomly divide the dataset into Train/Val/Test sets with 80/10/20 percent of the data in each. For the train and validation set, patches were generated as follows: within tissue areas, the high-resolution WSI was split into non-overlapping patches of 224x224 pixels at 10x magnification. Furthermore, to represent all the artefacts present in the dataset, patches were chosen at random in non-tissue areas totalling approximately half the training dataset. In total 240,181 patches were created, each with a corresponding GT mask. The 10x magnification was chosen as a compromise showing both the macro/micro-architecture of the tissue, as well as reducing the number of patches for storage and computation purposes.

**Testing set** 107 Test set WSIs were extracted at magnification 10X. Contrary to the Train/Val set the whole image was patched and reconstructed within the testing pipeline. Segmentation masks were predicted for the whole image and coloured as follows: Yellow for True Positive (TP) segmented tissue, Red for False Negative (FN) and Green for False Positive (FP).

**Training Schedule** Training was conducted using a UNet segmentation architecture as in [16]. All the models were trained using Adam optimizer with their default beta values; the learning rate was set to 0.0001 with a batch size of 20. A custom early stopping mechanism was employed to terminate the training before the model overfits the data. The loss and Dice score were monitored, and training halted if neither improved for five consecutive epochs. Data augmentation was used to modify Saturation, Hue, Contrast and Brightness values randomly. All networks were trained with the Focal Tversky loss function [1]. The Dice score metric was used to evaluate all results.

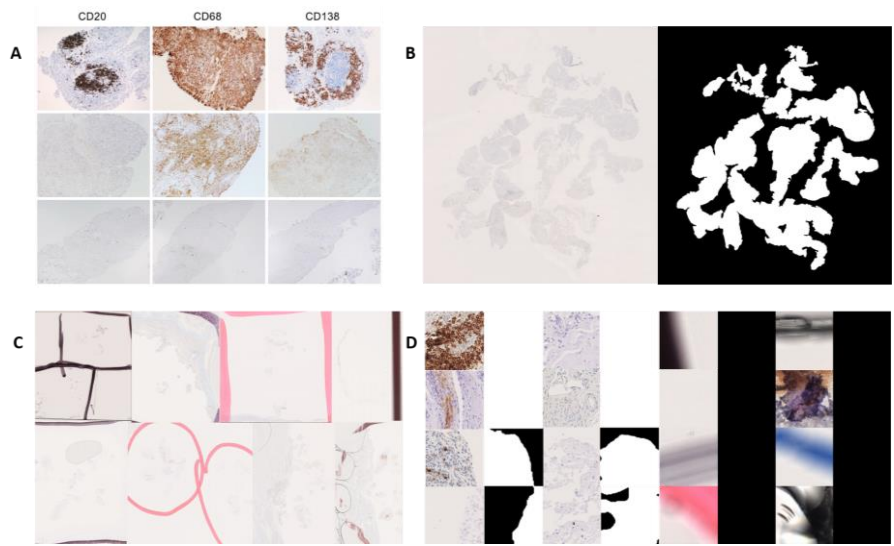

**Fig.1. A:** Immunohistochemistry of synovial biopsies for CD20+ B cells, CD68+ macrophages and CD138+ plasma cells from patients with early rheumatoid arthritis [9], showing the heterogeneity in staining colour and intensity. **B:** On the left, the original image, and on the right the hand-labelled ground truth binary mask. **C:** Examples of artefacts present in WSIs, such as pen marks, water drops, variable stain intensity, low contrast, blurriness, broken slide, etc. **D:** Examples of training images and their ground truth binary masks. Left: tissue areas. Right: artefacts.

## 3   Experimental Results

The UNet algorithm successfully segments synovial tissue WSIs stained with three different IHC stains and scanned in 20 different clinical centres across Europe obtaining a Dice score of $0.863 \pm 0.112$, indicating it is highly robust to different IHC stain types, clinical centre batch effect and data heterogeneity. In Figure 2, we show qualitative results to illustrate some of its strengths and limitations: in A2 we see the model is able to segment areas of low contrast, whilst ignoring strong signals such as pen marks. In B2, we see the model was actually more successful than the human labeller at correctly segmenting tissue areas, highlighting the need for an automated segmentation algorithm which does not suffer from fatigue or moments of inattention. Finally, in C2 we illustrate some of the limitations of the model, such as incorrect segmentation of speckled dye and iridescence stains. Further training of image artefacts could help improve robustness, yet overall the model is able to deal with common WSIs artefacts, such as variations in colour and intensity, blurriness, different colour pen marks, tissue folding, water drops, etc.

## 4   Conclusion

We present a fully automated deep learning segmentation algorithm, which is freely available and can be used as a first step in any rheumatoid arthritis or with further finetuning, any musculoskeletal IHC image analysis pipeline, avoiding lengthy manual annotation and helping to improve their speed, repeatability and robustness. This is a key step towards the acceleration of research into the mechanisms involved in rheumatoid arthritis and potentially other forms of inflammatory arthritis.

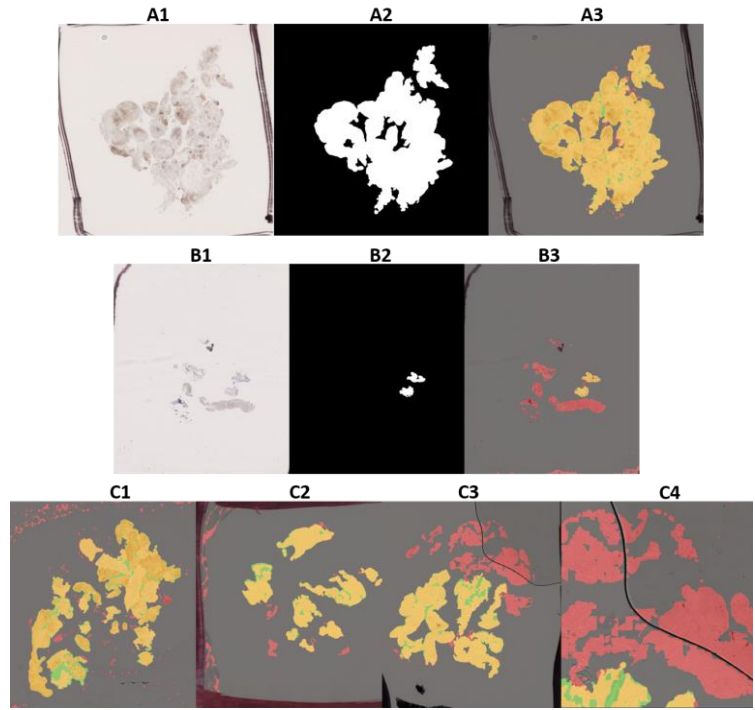

**Fig.2. A:** Example of a high Dice score results (0.97), with original image, hand-labelled segmentation mask and predicted segmentation mask. **B:** Example of a low Dice score result (0.20), with a poorly hand-labelled ground truth mask. **C:** in **C1** speckled dye spread on the slide is recognised as tissue. **C2** iridescent stain on the slide is recognised as tissue. **C3** a large area of tissue is classified in red as False Positive, however, this area corresponds to real tissue which was not annotated fully. In **C4** the algorithm correctly avoids segmenting the borders of a water drop.